\documentclass[12pt]{iopart}
\usepackage{graphicx}
\usepackage{iopams}
\begin{document}

\title[]{Resonant Ion Confinement  Fusion Concept}

\author{Jose Luis Rosales $^1$ and Francisco Castej\'{o}n $^2$}

\address{$^1$Centre for Computational Simulation  \\
 DLSIIS ETS Ingenieros Inform\'{a}ticos, Universidad Polit\'{e}cnica de Madrid,\\
Campus Montegancedo, E28660 Madrid, Spain.\\
 $^2$ Laboratorio Nacional de Fusi\'{o}n,  CIEMAT, Avenida Complutense 40
E28040 Madrid Spain.}
\ead{jose.rosales@fi.upm.es}
\vspace{10pt}

\begin{abstract}
Based on the theorized possibilities of resonant ion confinement, for a Deuteron cloud in a Penning-Malmberg trap with a specially configured rotating wall, the opportunity to design a new type of fusion device  is prospected. It is proven that, for some trap configurations, nuclear fusion reactions should take place and, in that case, Lawson's criterion for an efficient fusion reactor is met. Furthermore, the reactor could have a compact design and, since  it should not require a large facility, it can function as a fusion cell with a pure ion thermal gas.

\end{abstract}

%
\vspace{2pc}

\section{Introduction}
For an efficient fusion reactor, the pressure \small $p\;$\normalsize of the confined plasma times the energy confinement time
\small$\tau_E\;$\normalsize should be greater than a given amount
\small
 $ p\tau_E> L; \;$
\normalsize
which was a criterium due to Lawson (Ref.~\cite{Lawson}). For Deuterium - Deuterium reactions this value is of the order of one hundred \small $atm\times s.\;$\normalsize  This reaction happens in the stars, where plasma is confined by gravity. 
The advantage of considering Deuterium-Deuterium reactions is that this is a stable and abundant isotope of Hydrogen on Earth, although the cross section of these reactions is smaller than that of Deuterium-Tritium. Up to date, the magnetic confinement of neutral plasmas has been thought as the most promising concept to build a fusion reactor and two main concepts of devices have been developed: tokamaks and stellarators. In tokamaks, the magnetic field confines the particles and energy long enough for ignition to occur and it is expected that such a reactor will generate enough energy to achieve \small $Q> 10,\;$ \normalsize which will probably be shown through the ITER experiment in the near future (see Ref.~\cite{Donne}, also Ref.~\cite{L-G}). Nevertheless, tokamaks still have to overcome some challenges, namely the possibility of suffering disruptions, the existence of edge localised modes (ELMs) and the intrinsic pulsed working, on top of the possible impurity accumulation, (as from Ref.~\cite{Nature2021}). Turbulence will be also playing a major role degrading the Tokamak plasma confinement. These problems made it necessary to explore other possible magnetic confinement fusion concepts like the stellarator one (see e.g. Ref.~\cite{W7X}), characterised by enabling a continuous working and by the absence of large ELMs and disruptions, since there plasmas are almost currentless. Stellarator confinement is one generation after the tokamak one, due to the fact that they have to reduce their neoclassical transport by optimizing the magnetic configuration (see e.g. Ref.~\cite{Castejon-1} and references therein),as well as to solve the confinement of fast particles and to demonstrate power exhaust by a suitable divertor concept (see again Ref.~\cite{W7X}). Therefore, it is still necessary to overcome difficulties in both concepts to make it economically feasible for the production of electricity by fusion. Other topic to discuss in the future would be the size of the reactors, since all the scalings show an improvement of confinement with the size of the device (see e.g. Ref.~\cite{Castejon-2} and references therein). Thus, to address the problems encountered in the case of magnetically confined neutral plasmas and, at the same time, to explore alternative confinement methods that minimise the resources involved, another possibility to achieve nuclear fusion is described in this work; the new method is based on the concept of resonant ion confinement (see Ref.~\cite{patente}, which can be produced inside a cylindrical Penning-Malmberg ion trap with a Deuteron plasma in perfect thermal equilibrium). A remarkable feature of this fusion concept is the possibility of building small reactors, which brings with it completely different engineering challenges from those encountered in the design of magnetic confinement reactors.  Indeed, as we shall see, on the grounds of these resonant ionic confinement fundamentals, a kind of compact reactor, say a fusion cell,  can be conceived with, predictably, energy yields which, interestingly, are proportionally lower than those expected in ITER. These  fusion cell have, therefore, complementary purposes and scopes.

In this device ions are confined by applying an electric potential difference,\small $V_0,\;$\normalsize and a magnetic axial field \small $\mathbf{B}.\;$\normalsize The ion trap, additionally, has a quadrupole electric field of a frequency, \small$\omega,\;$\normalsize such that it allows the Deuterium nuclei to rotate also with that same angular velocity. In this system, a fraction of ions performs trajectories that should bring them close enough to produce fusion reactions by tunnelling. The confined ion cloud takes the shape of a highly flattened spheroid. Although these characteristics are common to any Penning-Malmberg configuration, there is a very precise set of values for the trap parameters that determines the intensity of the quadrupole field for which resonant, coalescing paths, arise among all the Deuterium ion trajectories, and, in this way, nuclear fusion reactions will likely take place.
Thus, the precise dependence of this quadrupole electric field as a function of the electric field potential, \small
$V_0,\;$ \normalsize and of the magnetic field intensity, \small $\mathbf{B},\;$\normalsize as well as the radius of the trap cylinder, \small$R_0,\;$\normalsize
will determine the resonance conditions required for the equipment to produce a finite number of fusion reactions. Finally, the pressure and confinement time necessary for ignition in the centre of the ionic trap should be reached only for special configurations of the
confinement system parameters.  What propose here, contrarily to other possible candidates for fusion devices, is to build a device that confines non-neutral plasmas in thermal equilibrium where, also, the ergodic theorem for the average trajectories of the confined ions is satisfied. 
\section{Methods}
\subsection*{Resonant Ion Confinement.}
Let us consider \small $N\;$\normalsize Deuteron nuclei of mass \small $m\;$\normalsize confined in a cylindrical
Penning-Malmberg trap of radius \small $R_0;\;$\normalsize we will assume perfect thermodynamic  equilibrium.  Let us denote the applied electrostatic
potential as \small $V_0\;$\normalsize and the axial magnetic intensity as \small$\mathbf{B}.\;$\normalsize  A rotating wall electric quadrupolar field of
intensity \small$\lambda=1/2+\delta\;$\normalsize and angular frequency \small$\omega\;$\normalsize  is then added which prompts the plasma cloud to rotate collectively with the same rotating wall angular frequency $\small \omega.\;$\normalsize The latter is the guiding orbit magnetron
frequency of the ion cloud. The charged plasma remains confined in the central plane of the trap and the form of the confined ion cloud is a
spheroid of semi-major axis \small $R_{C}\;$\normalsize and semi-minor axis \small$z.\;$\normalsize Using the fact that ions should be in
equilibrium inside the confined cloud, it is easy to see that the electric axial oscillations of ions should be related to the volume
density number,\small $n,\;$\normalsize of this cloud as $\small\omega_z=(n e^2/m \epsilon_0)^{1/2}.\;$\normalsize Moreover, the oscillator axial frequency also depends on the electric potential \small$V_0\;$\normalsize and the radius of the trap cylinder, \small $R_0,\;$\normalsize as \small
$\omega_z=(2/R_0)\sqrt{eV_0/m}.\;$\normalsize  In addition, ought to the applied axial magnetic field \small $\mathbf{B},\;$\normalsize the ions inside the
confined spheroid have an internal fast cyclotron oscillation whose frequency is defined in terms of \small$\mathbf{\Omega}=e\mathbf{B}/m.\;$\normalsize Due to the diamagnetic behaviour of charge currents inside the plasma cloud,  the achievable cyclotron frequency of the Deuterons is
actually smaller: \small $\Omega'=\Omega -2\omega,\;$\normalsize which this is called the vortex frequency. On the other hand, dynamic equilibrium of the plasma imposes that \small $\omega(\Omega-\omega)\rightarrow\omega_z^2/2.\;$\normalsize  This means that there exists a key parameter \small $\vartheta\;$\normalsize to define confinement quality of the Penning-Malmberg trap such that
\small$\omega=\Omega\sin^2\frac{\vartheta}{2}\;$\normalsize and \small $\omega_z=\Omega\sin\vartheta /\sqrt{2}.\;$\normalsize When \small $\omega_z\ll
\Omega\;$\normalsize the aspect ratio satisfies the condition \small $\ell=z/R_{C}\ll 1\;$\normalsize (see Ref.~
~\cite{Bollinger}). Ion confinement is expected to be perfectly stable free of the magnetohydrodynamic instabilities that happen in magnetic neutral plasma confinement. This is the confinement theorem of charged plasmas. 
In special situations, there are other kind of periodic solutions of the equations of motion (see Appendix). These solutions represent the trajectories of two correlated confined nuclei that follow coalescing paths to the centre of the trap, in the end this will  allow for their respectively trajectories to cross where they collide, and so D-D fusion reactions ought to happen with high probability through quantum tunnelling the Coulomb barrier. Yet, this effect will hold only at some given rate, since most of the ions in the cloud will still follow circular magnetron guiding orbits. Thus, the correlation only occurs for the special condition between the quadrupole field strength parameter \small
$\delta\;$\normalsize and the trap parameter \small $\vartheta\;$\normalsize takes place, which is the ion resonant confinement condition. To see this, recall that, in the rotating centre of mass frame of every two ions, the trap field interaction Lagrangian is given in terms of the  relative distance \small $\varrho\;$\normalsize  between the correlated nuclei as
\small
$
  L = \frac{1}{2}\mu\dot{\varrho}^2-U(\varrho,t),\;
$
\normalsize
the potential energy being \small $U(\varrho,t)=\frac{1}{2}\mu\omega_z^2 \varrho^2 (\lambda\cos2\omega t-\frac{1}{2}),\;$\normalsize where \small $\mu=m/2 \;$\normalsize stands for the reduced mass of the ions. It is seen that the fast cyclotron motion frequency does not appears in \small $U(\varrho,t).\;$\normalsize The solution of the equation of motion of the two-correlated-nuclei complex is given in terms of Mathieu functions. If \small $R_{C}\;$\normalsize is the maximum  radius  of the orbit, denoting,
\small
$\chi=\cot^2(\vartheta/2)\rightarrow \frac{eB^2 R_0^2}{4\mu V_0}-2\;$\normalsize
(for \small $\vartheta\ll 1$ \normalsize), one gets,
\small
$
\varrho=\frac{2R_{C}}{q_0} \bi{C}_e[-\tiny \chi,-\lambda\chi,\omega t\;],\;
$
\normalsize
where $q_0\;$\normalsize is a normalization constant. In general, these orbits are exponentially unstable, yet numerical calculations show that there exists a special, resonant,\small  $\pi/\omega\;$\normalsize periodic orbit whenever the quadrupolar field strength \small $\delta$
\normalsize takes the value, (see Appendix)
\small
\begin{eqnarray}\label{deltar}
 \delta\equiv\lambda-\frac{1}{2} \rightarrow  \frac{1}{\sqrt{2\chi}} .
\end{eqnarray}
\normalsize
The minimum distance between the nuclei in these resonant orbits will eventually be given by the following numerical solution
\small
\begin{equation}\label{min_r}
 \varrho_0=2 R_{C}\kappa , \;$\normalsize with \small $\; \kappa \simeq \exp\{1/2-1.35\sqrt{\chi}\},
\end{equation}
\normalsize
which can reach arbitrarily small values at any temperature of the confined plasma in the limit \small $\vartheta\rightarrow 0.\;$\normalsize
Moreover, the kinetic energy per Deuteron in this case becomes $E\rightarrow W/2$ where
\small
$
  W=\frac{e^2}{4\pi\epsilon_0 \varrho_0},\;
$
\normalsize
which provides \small $\varrho_0.\;$\normalsize 
On the other hand, a charged plasma in thermodynamic equilibrium at temperature \small $T\;$\normalsize
should behave as a solid rotor in a Penning-Malmberg trap with a magnetron frequency oscillatory  stroboscopic rotating wall quadrupolar field (see Appendix). The plasma global angular velocity \small $\omega\;$\normalsize, coinciding with that of  the individual ions guiding orbits, satisfies the condition \small  $\omega= 1/R_C\sqrt{(k_B T/\mu)}\;$\normalsize. Then, since \small $\omega= \Omega(\chi+1)^{-1},\;$\normalsize we get \small $\chi+1=\Omega R_{C}\sqrt{k_B T/\mu}\;$\normalsize which, owing to the dynamic internal equilibrium of the plasma, it is equivalent to saying that \small $\chi\simeq e V_0/(k_B T) \times (R_{C}/R_0)^2.\;$\normalsize Finally, using this thermodynamic constraint and given that the axial degree of freedom is thermal, i.e., \small $k_B T=1/2\mu \omega_z^2 z_{max}^2,\;$\normalsize we conclude that the aspect ratio of the confined ion cloud minimal height spheroid is \small $z_{max}/R_{C}\simeq 1/\sqrt{\chi},\;$\normalsize which is an important relation that will be used later. Moreover the probability that two correlated ions follow the resonant orbits is (see Appendix)
\small
$
\wp(\chi)=\chi^{-1}.\;
$
\normalsize
Recall that in the resonant case, all the macroscopic features of the Deuteron plasma can be written only in terms of the microscopic parameter $W$ and of the trap confinement value for \small $\chi.\;$\normalsize Then, \small $R_{C}=e^2/(8\pi\epsilon_0) W^{-1}
\kappa ^{-1}\;$\normalsize and the confined volume of the ion cloud becomes
\small
$ V\rightarrow
  \frac{4}{3}\pi (R_{C}/\sqrt{\chi})^3\times \chi,\;$
\normalsize
which is coincident with the volume of $\chi$  ion clump spheres of  radius \small $R_{C}/\sqrt{\chi}.\;$\normalsize
The density number becomes
\small
$
n=\frac{\epsilon_0 B^2}{4\mu}\sin^2\vartheta\rightarrow4\omega^2\epsilon_0 \mu\chi/e^2\; 
$
\normalsize
and the total number of confined Deuterons
\small
$  N\rightarrow V\times n.\;$ \normalsize

\section{Results}
\subsection*{Fusion Conditions.}
For Deuterium fusion, the occurring reactions  are
\small
 $^2H+^2 H  \to ^3 He+ n; \; and \;\;  ^2 H+^2 H  \to\;\; ^3 H\; +^1 H\;$
\normalsize
while their corresponding fusion energies are \small $E_{f1} [^3 He+ n] = 3.27 \;MeV$ \normalsize and \small $E_{f2} [^3 H\;
+^1 H] = 4.04\; MeV$\normalsize. Each reaction takes place with approximately \small $50\%\;$\normalsize probability. Thus, together with the necessary Lawson criterium, in a self-sustaining energy balanced fusion Deuterium device the  only heating terms shall be that from the kinetic energy of the confined charged fusion reactions products. The kinetic energy contribution of the neutrons, which cannot be confined in the trap, must be subtracted from the nuclear energy balance, a fraction \small $ \eta_n = E_n / (E_{f1} + E_{f2}).\; $\normalsize The neutron energy escapes directly to hit the reactor walls, where its energy can be extracted by a breeding blanket. On top of that, other energy losses have to be taken into account in the power balance analysis, namely the Bremsstrahlung term due to accelerated charges in the plasma, as well as the transport term, which is assumed to be due to ion-ion Coulomb collisions. Positive power balance of the Deuteron plasma, then,
requires
\small
\begin{equation}\label{Lawso}
P_f (1-\eta_n)>P_{B}+P_{L},
\end{equation}
\normalsize
where \small $P_f\equiv J_f V\;$\small is the power due to fusion reactions that heats the whole plasma volume \small
$V,\;$\normalsize \small $P_B\;$\normalsize is the {\it Bremsstrahlung} term and \small$P_L\;$\normalsize  is the power loss caused by transport, assumed collisional. Again, let \small $E\;$\normalsize be the energy of the Deuteron nuclei in the resonant orbit (this correlates with  the effective temperature of the nuclei {\it inside} the density clusters) and let  \small $n'\;$\normalsize be the density number of the plasma in the resonant clusters, then the collision frequency is, (see Ref.~\cite{Freidberg})
\small
$\nu(\chi,E)=n'E^{-3/2}\frac{3 e^4}{16\pi^2 \epsilon_0^2m^{1/2}} \ln\Lambda(\chi,E).\;$
\normalsize
The Coulomb logarithm is \small $\ln\Lambda(\chi,E)=\ln\big\{12\pi\epsilon_0^{3/2} n'^{-1/2}e^{-3}E^{3/2}\big\}.\;$\normalsize
Now, we write \small $N'=\wp(\chi) N\;$\normalsize as the number of resonant nuclei in the density clusters; then \small $V'=\kappa  V\;$\normalsize
is the volume of the density clump because only one of the two major axis of the spheroid is compressed due to the resonant orbits of the nuclei in
it.  Then every ion swirls around the centre in a vortex from \small $r=R_C\;$\normalsize to \small $r=\varrho_0/2\;$\normalsize and, therefore, the effective volume that affects all the resonant orbits can be approximated as \small $V'=4/3\pi R_C\big\{\kappa R_C\big\}z,\;$\normalsize which  will be the volume available for fusion reactions in the centre of mass of the resonant complex.
Then, \small $n'=\wp(\chi)/\kappa\times  n.\;$\normalsize  Now, from these definitions, the reactor effective power per unit of volume becomes
\small
\begin{eqnarray}\label{Sf}
 J =J_f -\eta \bigg\{ n'^2\alpha_{B} E^{\frac{1}{2}} +\frac{N'}{V} E\nu(\chi,E)\bigg \}
\end{eqnarray}
\normalsize
where the fusion power density is
\small
$J_f=\bigg\{\frac{1}{2}E_{f_1}\langle\sigma v\rangle_1+\frac{1}{2}E_{f_2}\langle\sigma v\rangle_2\bigg\}
\frac{n'^2}{2}.\;$
\normalsize
Notice that in Eq.~(\ref{Sf}),  we consider the total volume of the plasma in the collision term, instead of that of the resonant cluster of ions; the reason is that we consider the most unfavourable case in which the collisional transport is large enough to drive the local inhomogeneities that appear in the resonant zone to the entire plasma volume. In Eq.~(\ref{Sf})  $\eta\approx 1$ because, neglecting the \small $^3 H$, $^1 p \;$ \normalsize and \small
$^3 He_2\;$\normalsize  concentrations, i.e., those of the nuclear reaction products, there is only a single particle species in the confined plasma. For long working period of the reactor, the concentrations of these species could not be negligible and should be taken into account to develop a possible exhaust device. The trap disconnection can be used for exhauteing these particles in any case.
In Eq.~(\ref{Sf}), \small $\alpha_B=1.4\; 10^{-40}\;(m_e/m_p)^{3/2}\;\;[W/m^3 K^{-1/2}]\; $ \normalsize 
and,  for $E(keV) <300$, we can make the following analytical  approximation for the cross-section of the reaction:
\small
$\langle\sigma v\rangle_i  = b_i {E}^{\beta_i} \exp\{c_i {E}^{\chi_i}\}$
$b_1=1.198\times10^{-18}$,$b_2=3.5501\times10^{-19}$,$\beta_1=-1.0759$, $\beta_2=-0.9462$, $c_1=-23.511$, $c_2=-22.04$, $\chi_1=0.29221$,
$\chi_2=0.2922$.
\normalsize
Using these values in Eq.~(\ref{Sf}) one sees that the Bremsstrahlung and the collisional transport terms have small effects on the power
balance.

\subsection*{Fusion Cells.}

\begin{center}
\begin{figure}[hbtp]
\centering
\includegraphics[scale=0.65]{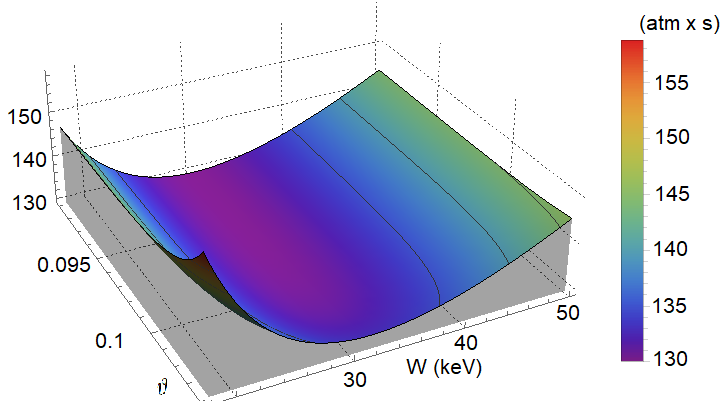}
\caption{Lawson triple product  \small $p\tau_E\;$\normalsize  as a function of the Coulomb barrier energy achievable for
D-D collisions \small $W\;$\normalsize in keV and the parameter \small $\vartheta=\arcsin(\frac{\sqrt{2}\omega_z}{\Omega}).\;$\normalsize As seen here, an stable minimum does exist, \small $p\tau_E\sim 90\;atm\times s,\;$\normalsize  for \small $\vartheta< 0.105\;$\normalsize and \small $W\sim 30.7 \;keV.$\normalsize}
\label{fig:Lawson1}
\end{figure}
\end{center}

\begin{figure}[hbtp]
\centering
\includegraphics[scale=0.65]{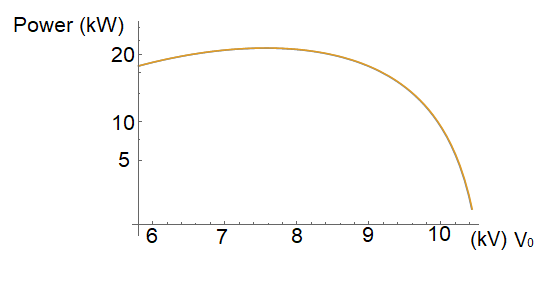}
\caption{ Optimal achievable power of the standard fusion device as a function of the applied confinement
electric potential \small $V_0,\;$\normalsize \small $B=3.73\; T,\;$\normalsize \small $W=30.7 keV\;$\normalsize and \small $\vartheta=0.0915.\; $\normalsize For the estimation, the frequency of the Deuteron refueling pulse was taken to be exactly coincident with the resonant one. }
\label{fig:Power}
\end{figure}

In order for the reactor to be energetically self-sustained, the generated energy must be greater than the sum of the emitted heat plus the magnetic back reaction pressure loss (as required from Brillouin's theorem). Then, the actual thermodynamically available power density is
\small
$ J_C= J-\dot{W},\;$
\normalsize
where \small $\dot{W}=\frac{b^2}{2\mu_0}\times \frac{\omega}{\pi}\;$ \normalsize is the power loss per unit of volume due to the diamagnetic currents of the confined ions  calculated for the resonance frequency. Recall that the contribution to the pressure losses is written in terms of the plasma back reaction magnetic field, namely,  \small $\mathbf{b}=-4(\mu/e)\mathbf{\omega}=-2  \mathbf{B} (1+\chi)^{-1}.\;$\normalsize
Now, the energy confinement time $\tau_E$ may be calculated in terms  of the temperature $T$ and the density number $n'$  for the nuclei in the resonance
 \small
$  \tau_{E}= n' E\times J_C^{-1}.\;$\normalsize
Since the pressure is \small  $p=2/3 n'E,\;$\normalsize then (see Ref.~\cite{Freidberg})
\small
\begin{eqnarray}\label{LawsonFusionCell}
 p\tau_{E}=\frac{2}{3}n'^2E^2/J_C.
\end{eqnarray}
\normalsize
The function $p\tau_E$ should have a minimum value for some \small $\chi\;$\normalsize and \small
$E=W/2.\;$\normalsize  As seen in Fig.~\ref{fig:Lawson1}, Lawson's minimum is actually reached for \small $\vartheta\leq 0.105,\;$\normalsize  and \small $W=30.7 \;keV.\;$\normalsize  Indeed, this minimum is obtained for approximately the same confinement parameters, independently of the actual temperature of the confined plasma cloud. Thus \small $p\tau_E\sim 90 \;atm\cdot s\;$\normalsize, which is expected to be experimentally achievable.  Moreover, the value of the Coulomb barrier energy for the Lawson minimum  computed  is remarkably close  to the expected experimental value (see Ref.~(\cite{Freidberg}), a fact that supports  the Resonant Ion Confinement reactor concept. Therefore, in order to obtain the actual power efficiency of the reactor, recall that, since the species \small$^3 He\;$, $p\;$\normalsize
and \small $^3 H\;$\normalsize are positively charged, they will  be retained in the trap but, contrarily to that, the neutrons will escape from the trap cavity and will be absorbed by the surrounding walls of the reactor chamber, where a separate equipment to absorb their energy should be installed. Their
kinetic energy might  be, then, transformed into heat (with an efficiency \small $\eta_h\sim 1/3\;$\normalsize).  This heat can used to produce electricity by means of high efficient thermoelectric materials.
Moreover, we must also take into account that every \small $D-D\;$\normalsize collision takes place by quantum tunnelling the barrier \small $W.\;$\normalsize  The Gamow-Sommerfeld  probability of this transition needed to undergo the nuclear reaction is 
\small
$
\eta_f (W)\sim\exp\{-\pi\alpha c\sqrt{2\mu/W}\}, \;
$
\normalsize
again, the Deuteron reduced mass is \small $\mu=m/2.\;$\normalsize It gives, for the Lawson minimum \small $W\simeq 30.7 \;keV,\;$\normalsize  \small
$\eta_f\simeq 1/300.\;$\normalsize   Having this in mind, the following estimate for the electric power of the reactor can be provided
\small
$\nonumber P_f\sim  \eta_n\eta_h J_C\times  V\rightarrow
\eta_n\eta_h J_C\times\{ \mathcal{Q}_tV'\},\;
$
\normalsize
where we have used the fact that, for the reactants in the cell, their actual available volume, \small $ V,\;$\normalsize must be calculated considering that the number of reactions in it must coincide, in average, with the sum of  the "\textit{Quantum tunneling}" configurations, \small $\mathcal{Q}_t$\normalsize,  all over the time, at some given space point; if this particular point is selected as the centre of the cell, we assume that \small $ V'=\kappa  V $\normalsize. Moreover, the quantity \small $\mathcal{Q}_t\;$\normalsize can be estimated as  \small $\mathcal{Q}_t=\eta_f(W)\times \wp(\chi)^2\times N^2/2,\;$\normalsize   in terms of the total number of available resonant ion pairs configurations times the tunnelling through the barrier probability. This taken into account, the following equation determines the key constraints for the reactor confinement configurations
\small
\begin{equation}\label{ergodic}
  \mathcal{Q}_t \times \kappa  = 1,
\end{equation}
\normalsize
for \small
$R_C\leq R_0,\;$
\normalsize
where the radius of the Penning-Malmberg trap should be \small $R_0=2(\chi+1)(eB)^{-1}\times (eV_0\mu/\chi)^{1/2}\;$\normalsize and the maximum radius of the confined plasma cloud  \small $R_C=\frac{1}{2}\varrho_0\kappa ^{-1}\;$\normalsize.  The total thermoelectric power  becomes
\small
\begin{equation}\label{Thermolectric_Power}
 \mathbf{P}=P_f-(1-\eta_h)a T^4[\pi/\omega] V.
\end{equation}
\normalsize
Here, to obtain the realistic model of the fusion device, we have subtracted the black body radiation term ($a$ is the radiation constant), recall also that \small  $k_B T\simeq eV_0/\chi \times (R_C/R_0)^2.\;$\normalsize In addition to that, the maximum variation of the number of Deuterons per unit of time in the reactor is \small $  \dot{N}_D=\frac{1}{\eta_n}
P_f/\langle E_f\rangle\; $ \normalsize and, therefore, the thermonuclear reaction frequency is  $\nu_f=\dot{N}_D/N$. On the other hand, the D-D collision frequency should be of the order of the cyclotron one, \small $ \Omega.\;$\normalsize Thus, for the actual resonant ion confinement device, we propose that the Deuteron  fuel will enter the resonant confinement device cavity in a pulsed way. To this point, the fuel is introduced in the reactor chamber through very high frequency pulses and the achievable electric power of the ion cell will depend on the ultra high frequency and high voltage circuit relays to control the rapid pulsed fuel refilling of the reactor chamber. Then, if we denote the frequency of these circuit relays as  \small $ \nu_{CB}\;$\normalsize, the achievable power would be: \small$2\pi\nu_{CB}/\Omega \times \mathbf{P}.\;$\normalsize
A solution of the ergodic constraint in Eq.~\ref{ergodic} does exist and is \small $\chi =477.102,\;$\normalsize \small $B=3.73\;
T,\;$\normalsize  and \small$W= 30.74 \;keV.\;$\normalsize  Recall that, remarkably,  the  solution of the ergodic condition for the barrier energy \small $W\;$\normalsize  coincides with that provided by the stable Lawson point from the nuclear model alone. The resulting maximum achievable electric power would be
\small
$
\mathbf{P} \sim 11\times 2\pi\nu_{CB}/\Omega\; (MW)
$
\normalsize
for Deuteron pulses of \small $N=2.2\times 10^{10}\;$\normalsize ions, if the electric potential is within the range \small $5.9
\;kV<V_0<10.5\;kV.\;$\normalsize In order to estimate the actual achievable power of the cell a possible  approximation should be to
provide \small $2\pi\nu_{CB}< O(\omega),\;$\normalsize  i.e, that, in order to allow for the pulses to reach the thermal
equilibrium during confinement, and to the aim to achieve the resonance,  the refueling frequency  should be slightly smaller than the resonant one.  In that case, a solution of  Eq.~\ref{ergodic}  is represented in Fig.\ref{fig:Power} for the attainable thermoelectric power in
Eq.~\ref{Thermolectric_Power}. This configuration gets a  maximum  of approximately
\small
$$
\mathbf{P} \sim 20\; kW\;
$$
\normalsize
at \small $V_0\simeq 8 \;kV\;$\normalsize for a fuel rate of $\dot{N}_D\simeq 2.1\times 10^{17}\;Deuteron/s.\;$ The confinement radius of
the ion cloud would be \small $R_C\simeq 91 \;mm\;$ and the trap radius is \small $R_0\simeq 92.56
\;mm.\;$\normalsize The actual fusion device requires a pulsed refilling of the reactor cavity and that the resonant configuration would be attainable only adiabatically. This kind of technological requirements will be the subject of the experimental research for real fusion facilities based on the grounds of the resonant ion confinement concept. Yet, in order to double check that the above estimates are correct, we can do a kind of Fermi analysis. As said,  in the resonant cell (all over the time), the
number of possible configurations between two Deuterons, leading to quantum tunneling reactions, is approximately \small $\mathcal{Q}_t\sim\eta_f(W)\wp
(\chi)^2\times N^2/2.\;$\normalsize Since the natural colliding frequency is \small $\Omega/\pi,\;$\normalsize the nuclear reaction rate should be
\small$\mathcal{Q}_t\times \Omega/\pi,\;$\normalsize and the number of emitted neutrons per second is half this value. Then, roughly, the actual
achievable electric power must be \small $\mathbf{P}\sim\eta_h 2.44 \;MeV \times [\frac{1}{2}\mathcal{Q}_t\times \Omega/\pi]\times
(\omega/\Omega)\;$\normalsize (the last term takes into consideration that we refuel the cell at the typical resonant frequency rates). For the numbers above, i.e., \small$B=3.7\;T,\;$\normalsize \small$N=2\times
10^{10}\;Deuterons,\;$\normalsize \small$\chi\sim 477,\;$\normalsize and \small $R_C\sim 91 \;mm \;$\normalsize (which, for the resonant
condition in Eqs.~\ref{min_r}, is equivalent to saying that \small$\varrho_0\simeq 47 \;fm$ or  \small $W\simeq 30.7\; keV\;$\normalsize),
we get \small $\mathbf{P}\sim 26\; kW,\;$\normalsize which is of the order of the figure that we did obtain from nuclear theory
alone. This fact makes us confident in the correctness of the derived result. We can call this optimum and small size configuration of the confinement device parameters attaining the maximum power as a fusion cell.

\section{Discussion}

For a compact reactor of small size, a new type of fusion technology can be developed according to the basic conditions described in this article, although some practical considerations must be taken into account. This possibility arose from imposing the parametric relations in Eqs.~\ref{deltar} and~\ref{min_r}  to the standard Penning-Malmberg ion trap, which lead to the necessary resonant kind of ion confinement. On the other hand, the trap fusion conditions are the key configurations that satisfy the  Eq.~\ref{ergodic} and, very promisingly, both the resonant and the operative conditions for Deuteron confinement are found within the range of the current state of the art of technological capabilities. In other words, due to the resonance, the Coulomb barrier can be overcome due to the movement of resonant ions. Additionally, one must expect that a pure Helium-3 resonant ion confinement device will work as another  fusion  possibility in exactly the same principles as the Deuterium one exposed in this article. Recall that such Helium-3 Cell will be a neutronic one and, as a safe reactor, it opens the possibility to implement a sort of thermonuclear battery in the event that it could power an autonomous device provided with some finite reservoir of Helium-3 gas. Even though the physics of both types of cells should be very alike, it does not necessarily imply that their engineering designs, which have to be associated with solving the problems of each type of reactors, should be also, in turn, similar. Therefore the Helium-3 cell must be studied separately and the results will be presented elsewhere.

\section{Conclusion}

In this work, the theory of resonant ion confinement has been developed for a charged gas in thermal and dynamic equilibrium.  Precise relationships between the trap parameters and Deuteron-Deuteron collision probabilities have been developed for ions following coalescent trajectories, leading, this way, to fusion reactions. The analysis takes into account the statistical ergodicity criterion of the system, as well as the assumption of thermal equilibrium of the ion gas, which should be necessary, we claim, for any sustainable and stable fusion reactor. In these circumstances, the Lawson criterion is fulfilled.  This device is capable of producing $20$ kW of electrical power with a small reactor design that we have termed a fusion cell.  Despite of the final energy production efficiency, the experimental test of this concept must be based on the eventual emitted neutron measurements.

\vspace{3 mm}

{\it Acknowledgements.} Jose Luis Rosales has been supported by an internal research contract owned by the Universidad Polit\'{e}cnica de Madrid, he is also indebted to previous collaborations with Professor Vicente Mart\'{i}n leading to the concept of Resonant Ion Confinement in Penning-Malmberg traps.

\appendix

\setcounter{section}{-1}

\section{}

A cylindric  Penning Malmberg trap is  characterised by the introduction of an axial magnetic field \small $\mathbf{B}\;$\normalsize and a electrostatic potential \small $V_0,\;$\normalsize in a cylindrical cavity where the ions are introduced. Additionally the full stability can be obtained with the help of a rotating electric field with  angular frequency \small $\omega$ \normalsize (see Ref. \cite{Bollinger}) and field strength \small $\lambda;\;$\normalsize  in this case, the motion of the ions in the trap is decomposed into separated rotationally confined mode and  an axial oscillation.  Let \small $q\;$\normalsize be the charge of the ion, then, the Lagrangian  is
given in terms of the electrostatic quadrupole and the magnetic field frequencies of the trap, i.e., denoting the cyclotron frequency \small $\Omega=qB/m$\normalsize
\small
\begin{equation}\label{L}
 L=\sum_i\frac{1}{2} m\{\dot{x}_i^2+\dot{y}_i^2 +\dot{z}_i^2+\Omega(x_i\dot{y}_i-y_i\dot{x}_i)\}
 - U_t(\mathbf{r}_i),
\end{equation}
\normalsize
\small $U_t(\mathbf{r})= q V_0+ \frac{1}{2} m \{\omega_z ^2 [z^2-\frac{1}{2}\rho^2]+\lambda\omega_z ^2[ (x^2-y^2)\cos 2\omega t-2 xy \sin2\omega t]\},\;$
and $\rho^2=x^2+y^2.\;$\normalsize This rotating quadrupolar electrical
potential wall of intensity $\lambda$ might be  required for the adiabatic stability of the
ions in the trap as it will be confirmed  down from.  Let us consider the rotating coordinate frame which rotates with angular frequency \small $\omega$: $(x,y)\rightarrow (\xi,\zeta).\;$\normalsize
Then  the rotating wall quadrupole perturbation becomes time independent,
\small
$(x^2-y^2)\cos(2\omega t) -2 xy \sin(2\omega t)\rightarrow \xi^2-\zeta^2,\;$
\begin{equation} \label{L1}
L=\sum_{i}\frac{1}{2} m\{\dot{\xi}_i^2+\dot{\zeta}_i^2 +\dot{z}_i^2+(\Omega-2\omega)(\xi_i\dot{\zeta}_i-\zeta_i\dot{\xi}_i)\}-U'(\mathbf{r}_i),
\end{equation}
\normalsize
\small $U'(\mathbf{r})= q V_0 +\frac{1}{2} m\omega_z ^2 z^2 - \frac{m\rho^2}{4} \omega_z ^2+\frac{m\rho^2}{2} \Omega\omega-\frac{m\rho^2}{2} \omega^2+\frac{m}{2}  \lambda\omega_z ^2(\xi^2-\zeta^2).\;$\normalsize
Hereafter the constant energy potential term \small $q V_0 \;$\normalsize will be omitted. The \small $z\;$\normalsize coordinate motion is an harmonic oscillator of frequency $\omega_z$. The restoring force gives the effective axial component of the electric field \small $e E_z=-m\omega_z^2 z$\normalsize. Along these lines, if the density of the plasma cloud is $n$, \small$E_z\;$\normalsize can be thought as that coming from the displacement of a positive charge in the plasma cloud (effectively a relative negative charge)  \small $\Delta q=- q z\oint dS n$\normalsize. Then, \small $\oint E_z dS=\Delta q /\epsilon_0\;$\normalsize  and from this the charge density becomes
\small
$  n=\frac{m\omega_z^2\epsilon_0}{q^2}.\; $
\normalsize
On the other hand, in order to compensate the axial displacement of every ion, this motion should be correlated with that of  some other ion in the opposite \small $z\;$\normalsize coordinate; moreover this z axis degree of freedom can be assumed to be thermal and, therefore, for these two correlated ions  \small
\begin{equation}\label{Temperature}
 2\times \frac{1}{2} k_B T=\frac{1}{2}\mu\omega_z^2 z_{max}^2.
\end{equation}
\normalsize
Then, in order to determine  the orbits of what may represent those  complementary bundles of  ions, let us  first recall that  in the rotating frame the net quadrupole  radial force can be averaged out to zero in every cycle (\small $\langle F_{q}\rangle = -\lambda m\omega_z^2\langle\rho(t)\cos2\omega t\rangle \to 0,\;$ \normalsize for \small $0\leq\omega t\leq  2\pi \;$\normalsize). Indeed, three forces are acting on each of the ions, namely, the centrifugal force \small$+m\omega^2 \rho\;,$ \normalsize the radial electric force \small $+\frac {1}{2}m\omega_z^2\rho\;$\normalsize and that one associated with the radially electric field induced by the rotation \small $\omega\;$\normalsize through the direction of the axial magnetic field whose value is \small $-m\omega\Omega\rho.\;$\normalsize It is this field that will  provide the radial confinement. To see how, recall that, from the symmetry of a cylindrical trap, the total angular momentum (that from the ions plus that from the magnetic field),  \small $\sum_{i}\{ m_i v_{\theta \; i}\rho_i +qB\rho_i^2/2\},\;$ \normalsize is preserved and, therefore, for instance for large $B$, \small $\sum_i \rho_i^2\; $\normalsize must reach some constant, which, as said, implies radial confinement; this argument is due to O'Neil (Ref. \cite{oneil}, Ref. \cite{Oneil1}).  With that in mind, the simplest case where the perturbed quadrupole frequency should be taken as that case where the net radial force acting on each of the charges is zero. Then,
\small
$
  \omega_z^2=2\omega(\Omega-\omega).\;
$
\normalsize
It means that  a "trap angle" can be defined such that
$
\small
\omega_z^2=\frac{1}{2}\Omega^2\sin^2 \vartheta,\;\; \omega=\Omega\sin^2(\frac{\vartheta}{2}).\;
$
\normalsize
This condition corresponds to the confinement situation of a thin, rigidly  rotating, spheroid of plasma at angular velocity $\omega$ (Ref. \cite{Dubin} Ref. \cite{Bollinger2}). Recall also that for that confined cloud of ions  of mass \small $m\;$ \normalsize and charge \small $q=Ze\;$\normalsize  a relation between the axial frequency \small $\omega_z\;$\normalsize, the cylindric radius \small $R_0\;$\normalsize and the applied electric potential  \small $V_0\;$ \normalsize can be found as the solution of \small $V(R_0)=0,\;$\normalsize where  \small $V(R)= V_0-\frac{m}{4}\omega_z^2 R^2/q.\;$\normalsize This gives,
\small
 $\omega_z^2=\frac{4q V_0}{m R_0^2},\;$
\normalsize
Lagrange's equations  read (denote \small $\tau(t)=\Omega' t\; $\normalsize and  \small $\varepsilon=\lambda \omega_z^2/\Omega'^2  $ )\normalsize
 \small
\begin{eqnarray}\label{LagrangeEquations}
  \frac{d^2\xi}{d\tau^2}-\frac{d\zeta}{d\tau}+\xi\varepsilon=0, \;\;\;
  \frac{d^2\zeta}{d\tau^2}+\frac{d\xi}{d\tau}-\zeta\varepsilon=0.
\end{eqnarray}
\normalsize
Elseways, since the Lagrangian is not time dependant, the energy is preserved and a first integral of motion is obtained \small $E=\textstyle \sum_{i}\{\dot{\xi_i}\partial_{\dot{\xi_i}}L+\dot{\zeta_i}\partial_{\dot{\zeta_i}}L\}-L=\textstyle \sum_{i}\frac{m}{2}\{\dot{\xi_i}^2+\dot{\zeta_i}^2+\lambda\omega_z^2(\xi^2-\zeta^2)\}.\;$\normalsize This means that the motion is bounded and that a time period $t_0$ can be found such that \small $ 2E/(\Omega'^2 m)=\varepsilon\textstyle \sum_{i}\{\xi_i(t_0)^2-\zeta_i(t_0)^2\}.\;$\normalsize
As said, the orbits of the  ions are combinations of  rapid  bare cyclotron  \small $\Omega'\;$\normalsize oscillations  plus a slow guiding centre magnetron trajectory of angular velocity \small $\omega\;$\normalsize which becomes stabilised by the rotating quadrupole force.  Notwithstanding with this, other solutions can be seen as representing  binary collisions at the centre of the trap. Now, define,\small $\;\eta=\sqrt{1+4\varepsilon^2},$ \normalsize  \small $\; \sigma =\sqrt{1/2(\eta-1)}\rightarrow  \varepsilon\;$\normalsize,  \small $ \gamma = \frac{\varepsilon+\sigma^2}{\sigma}\rightarrow  1\;$\normalsize, \small $ \beta = \sqrt{1/2(\eta+1)}\rightarrow 1\;$\normalsize and $\gamma'= -\frac{\beta}{\varepsilon+\beta^2}\rightarrow -1\;,$  \normalsize for \small $\varepsilon\ll 1.\;$\normalsize With this notation the exact solutions are
\begin{eqnarray}\label{Solutions}
\nonumber \xi_{h}=a\cosh\sigma\tau,\;\; \zeta_{h}=a\gamma\sinh\sigma\tau,\;\;\;\;\;\;\;\;\;\\
\xi_{C}= b\cos\beta\tau,\;\;\;\;\; \zeta_{C}=b\gamma'\sin\beta\tau.\;\;\;\;\;\;\;\;\;
 \end{eqnarray}
\normalsize
The $\xi_{C}$, $\zeta_{C}$ should correspond to the mentioned bare cyclotron orbits  (see e.g., Hasegawa  et al. Ref. \cite{Bollinger} for the solutions of the cyclotron orbits in the rotating wall trap configuration).
Thus, for instance, if the general solution is  \small $\vec{r}=\vec{r}_h+\vec{r}_C,\;$ \normalsize  the limit \small $\varepsilon\rightarrow 0\;$\normalsize is just the usual Penning trap constant radius magnetron orbit $|\vec{r}|=a$ provided with a rapid cyclotron oscillation  around this guiding orbit. Incidentally, in spite of the fact that $\xi_{h}$, $\zeta_{h}$ can be seen as spurious solutions  (they are not bounded hyperbolae that do not meet the required confinement conditions), recall that, admissibly, some of the actual orbits could also be represented by linear combinations of these hyperbolic plus the bare cyclotron solutions:
\small
  $\xi^{(i)} =\xi_{h}^{(i)}+\xi_{C}^{(i)}, \;\; \zeta^{(i)} =\zeta_{h}^{(i)}+\zeta_{C}^{(i)}.\; $
\normalsize
These trajectories would  exist during some period of time \small $t_0,\;$\normalsize say, \small $-t_0\leq t\leq t_0,\;$\normalsize. Numerical simulation shows that, in the rotating frame, the trajectory is just an hyperbola provided with rapid cyclotron oscillations having two turning points (at which \small $\dot{\xi}(t_0)=\dot{\zeta}(t_0)=0.)\;$\normalsize  Is obvious that the physical situation corresponds to the orbits of two long range coupled ion density fluctuations that collide (and repel each other) at the centre of the trap with \small $\xi^{(1)}=-\xi^{(2)}, \; \zeta^{(1)}=-\zeta^{(2)}.\;$\normalsize This is obviously correct because, otherwise, single ions would not preserve linear momentum individually.
The only relevant degree of freedom ought be the relative distance between the two charged aggregates, \small $\varrho =2(\xi^{2}+\zeta^{2})^{1/2}\;$\normalsize. For this coordinate, the collision is described from the Lagrangian  Eq.~\ref{L1} replacing \small $\xi=\varrho\cos\omega t,\;\;\zeta=\rho\cos\omega t,\;$\normalsize and \small $m\rightarrow\mu$\normalsize
\small
\begin{equation}\label{Effective_Mathieu_Lagrangian}
  \mathcal{L} = \frac{1}{2}\mu\{\dot{\varrho}^2+\frac{1}{2}\omega_z^2 \varrho^2-\omega_z^2 \varrho^2\lambda\cos 2\omega
  t\}.
\end{equation}
\normalsize
Denoting \small $\varphi=\omega t\;$\normalsize  and
\small
$
 \mathbf{\chi}=\frac{\omega_z^2}{2\omega^2}= \cot^2(\frac{\vartheta}{2})\;
$
\normalsize
 the equation of motion reads
\begin{equation}\label{Mathieu}
   \frac{d^2}{d\varphi^2}\varrho-\{\mathbf{\chi}-2\lambda\mathbf{\chi} \cos 2\varphi \}\varrho=0.
\end{equation}
Eq.~\ref{Mathieu} is Mathieu's equation whose time symmetric solution is obtained in terms of the Mathieu Cosine function:
\small
$\varrho(\varphi)= \frac{2R_C}{c}  C_e(-\mathbf{\chi},-\lambda\mathbf{\chi},\varphi),\;\;\;\;\; c=C_e(-\mathbf{\chi},-\lambda\mathbf{\chi},0).\;$
\normalsize
Thus, as in the case of the inverse pendulum, there are periodic stable bound solutions only within a very narrow parametric region \small $\lambda(\mathbf{\chi})\;$\normalsize (see~Ref. \cite{Meirov}), they have period  $\pi$ for the variable \small $\varphi.\;$\normalsize   For \small $\omega_z\gg\omega,\;$\normalsize numerically,  this parametric
stability constraint corresponds to a dependency between the quadrupolar electric force intensity and the parameters of the Penning trap
\small
\begin{equation}\label{resonant condition}
\lambda\rightarrow \frac{1}{2}+1/\sqrt{2\mathbf{\chi}}
\end{equation}
\normalsize
Additionally,  the closest distance between the two ion bundles defines the squeezing  factor, \small $\kappa ,\;$\normalsize of the coalescing orbits, i.e., numerically
\small
\begin{eqnarray}\label{squeezing_factor}
\ln\{\varrho_0/2R_C\}\rightarrow\ln \kappa \rightarrow  \frac{1}{2}-1.35\sqrt{\mathbf{\chi}}
\end{eqnarray}
\normalsize
here \small $\varrho_0\equiv\varrho(\pi/2)\;.$\normalsize The implication is that, if the resonant condition in Eq.~\ref{resonant condition}  is satisfied,  any closeness, even small, between the two positive ion clumps can be reached near the centre of the trap when \small $\mathbf{\chi}\gg 1.\;$\normalsize Eqs.~\ref{resonant condition} ~and~\ref{squeezing_factor} constitute the grounds of the Resonant Ionic Confinement Method. Owing to the existence of the quadrupole,  the magnetron degree of freedom is stabilised and the plasma is  macroscopically described as a rigid rotor of angular velocity \small $\omega.\;$\normalsize  Notwithstanding with this, microscopically, each  individual ion radial velocity should be Maxwellian and the stability of the plasma requires that the solid rotor energy be thermal. Incidentally, the actual inertia momentum of every "rigidly rotating" stabilised thin disk of  plasma becomes  \small $I_{Disk} =\sum_{i\in Disk} m_ir_i^2= \frac{1}{2} N_{Disk} mR_C^2,\;$\normalsize  providing a rotational energy giving by \small $E_{Disk}=\frac{1}{2}I_{Disk} \omega^2;\;$\normalsize yet,  to this rotational degree of freedom of each individual  ion,  a thermal energy \small$ k_B T/2\;$\normalsize  must be allocated. Along these lines,  the condition of thermal equilibrium  of the stabilised, rigidly rotating, plasma is compelled to be
\small
$
N_{Disk} \times\frac{ k_B T}{2}=\frac{1}{2}I_{Disk} \omega^2.\;
$
\normalsize
It imposes
\small
  $\omega R_C= \sqrt{ k_B T/\mu }\;$
\normalsize
which, since \small$\omega\simeq\Omega/\chi,\;$\normalsize 
also implies, in the approximation  \small$\omega_z\gg \omega,\;$\normalsize that \small   $\chi\sim \frac{qV_0}{k_B T}(\frac{R_C}{R_0})^2.\;$ \normalsize
Additionally,  the density clumps in the  coalescing orbits are practically confined in the centre of the trap most of the time and they move away to reach some maximum confinement radius \small $R_C\leq R_0\;$ \normalsize where they bounce back again to the centre. In this case, the ions inside those large aggregates, may interact individually when, owing to the resonance, the relative distance between the clumps is  reduced to a tiny minimum.    It is straightforward to derive, the following  relations for  \small $\chi\gg 1 :\;$\normalsize
\small
$
 \;B\simeq \frac{2\chi}{q R_C}\sqrt{ k_B T \mu},\;
 n\simeq \frac{\epsilon_0}{\mu\chi} B^2,\;
 z_{max}\simeq R_C/ \sqrt{\chi},\;
 N\simeq n\chi\times V_c,\;
$
\normalsize
where \small $  V_c= V/\chi=\frac{4}{3}\pi (R_C/\sqrt{\chi})^3\;$\normalsize is the clump volume.  Therefore, we see,  the confined plasma volume can be calculated as if there were \small $\chi\;$\normalsize granular spheres of ions of radius \small $ R_C/\sqrt{\chi}.\;$\normalsize From these equation it is possible to derive that
\small
$$
n= \{(N/\chi)\; 2\times\frac{1}{2} k_B T/E_{C}\} \times {(N/\chi) }/\frac{4}{3}\pi R'^3\}
$$
\normalsize
where \small $R'=R_C/\sqrt{\chi},\;$\normalsize whereas, \small $E_{C} =(qN/\chi )^2 /(8\pi\epsilon_o R'),\;$\normalsize corresponds to the Coulomb energy of a  bubble of \small $N/\chi\;$\normalsize ions on the surface of a  sphere whose radius is precisely \small$R'.\;$\normalsize Yet, recall that the plasma is fully thermal, imposing that the two dimensional surface energy  satisfies \small $2\times \frac{1}{2}(N/\chi) k_B T  = E_{C}.\;$\normalsize These thermal conditions of the plasma are, indeed, fully compatible with the existence of small fluctuations in the statistics that, according to our interpretation of the two coalescing ion density clumps, in the resonant case,  will orbit the cloud at periodic Mathieu trajectories. For some  clump pair in the plasma, the resonant situation will hold  with some probability, say \small$\mathbf{\wp}[i\in\{coalescing\}].\;$\normalsize  To estimate this recall that, owing to the periodic radial displacement of the density bundles, analogously to the axially periodic degree of freedom, dynamical equilibrium imposes that there should be some effective radial restoring force \small$q \delta E_\varrho=-\mu(2\omega)^2\delta \varrho.\;$\normalsize Again, this corresponds to a  negative effective displaced charge (a hole in the positively charged ion cloud) of \small $-q\delta \varrho n' \oint dS,\;$\normalsize where \small $ n'\;$\normalsize is the density of the displaced density clumps inside the plasma. Now, Gauss theorem applied to a thin disk surface of the plasma states that \small $ n'= 2m\epsilon_0\omega^2/q^2,\;$\normalsize which, from its statistical definition \small$ n'\equiv n \mathbf{\wp}, \;$\normalsize gives
\small
\begin{equation}\label{probability}
  \mathbf{\wp}=\frac{\omega}{\Omega-\omega}=\mathbf{\chi}^{-1},
\end{equation}
\normalsize
which, given that \small $\omega<\Omega/2,\;$\normalsize it is always lower than \small $1\;$\normalsize  as it should be.
Recall that near to the centre of the trap, the minimum  distance between the colliding ions can be approximated by
\small
$
  \varrho(t)\simeq |\varrho_0+ 2R_C\varepsilon\cos\Omega't|,\;
$
\normalsize
it means that the distance obtains its minimum when \small $\Omega'\Delta t=\pi.\;$

\end{document}